\documentclass{article}
\usepackage{spconf,amsmath,graphicx}

\usepackage{enumitem}
\setlist{nosep, leftmargin=14pt}

\usepackage{mwe} 

\usepackage{booktabs}
\usepackage{bm}
\usepackage[super]{nth}
\usepackage{amssymb}
\usepackage{hyperref}

\title{Clinically Relevant Latent Space Embedding of Cancer Histopathology Slides Through Variational Autoencoder Based Image Compression}

\name{
    Mohammad Sadegh Nasr$^{\star, 1, 2}$ \qquad
    Amir Hajighasemi$^{\star, 1, 2}$ \qquad
    Paul Koomey$^{1, 2}$
}

\secondlinename{
    Parisa Boodaghi Malidarreh$^{1, 2}$ \qquad
    Michael Robben$^{2, 4}$ \qquad
    Jillur Rahman Saurav$^{1, 2}$ \qquad
}

\thirdlinename{
    Helen H Shang$^{1, 3}$ \qquad
    Manfred Huber$^{1}$ \qquad
    Jacob M. Luber$^{\dagger, 1, 2, 4}$
}

\address{
    $^{1}$ Department of Computer Science and Engineering, University of Texas at Arlington\\
    $^{2}$ Multi-Interprofessional Center for Health Informatics, University of Texas at Arlington\\
    $^{3}$Division of Internal Medicine, Ronald Reagan University of California Los Angeles Medical Center\\
    $^{4}$ Department of Bioengineering, University of Texas at Arlington
}

\begin{document}
    
    \maketitle
    
    \def\thefootnote{$\star$}\footnotetext{These authors contributed equally to this work.}
    \def\thefootnote{$\dagger$}\footnotetext{Responsible author. Email: \texttt{jacob.luber@uta.edu}}
    
    \begin{abstract}
    In this paper, we introduce a Variational Autoencoder (VAE) based training approach that can compress and decompress cancer pathology slides at a compression ratio of 1:512, which is better than the previously reported state of the art (SOTA) in the literature, while still maintaining accuracy in clinical validation tasks. The compression approach was tested on more common computer vision datasets such as CIFAR10, and we explore which image characteristics enable this compression ratio on cancer imaging data but not generic images. We generate and visualize embeddings from the compressed latent space and demonstrate how they are useful for clinical interpretation of data, and how in the future such latent embeddings can be used to accelerate search of clinical imaging data.

\end{abstract}


    \section{Introduction}
\label{sec:intro}

    \begin{figure}[ht!]
        \centering
        \centerline{\includegraphics[width=\linewidth]{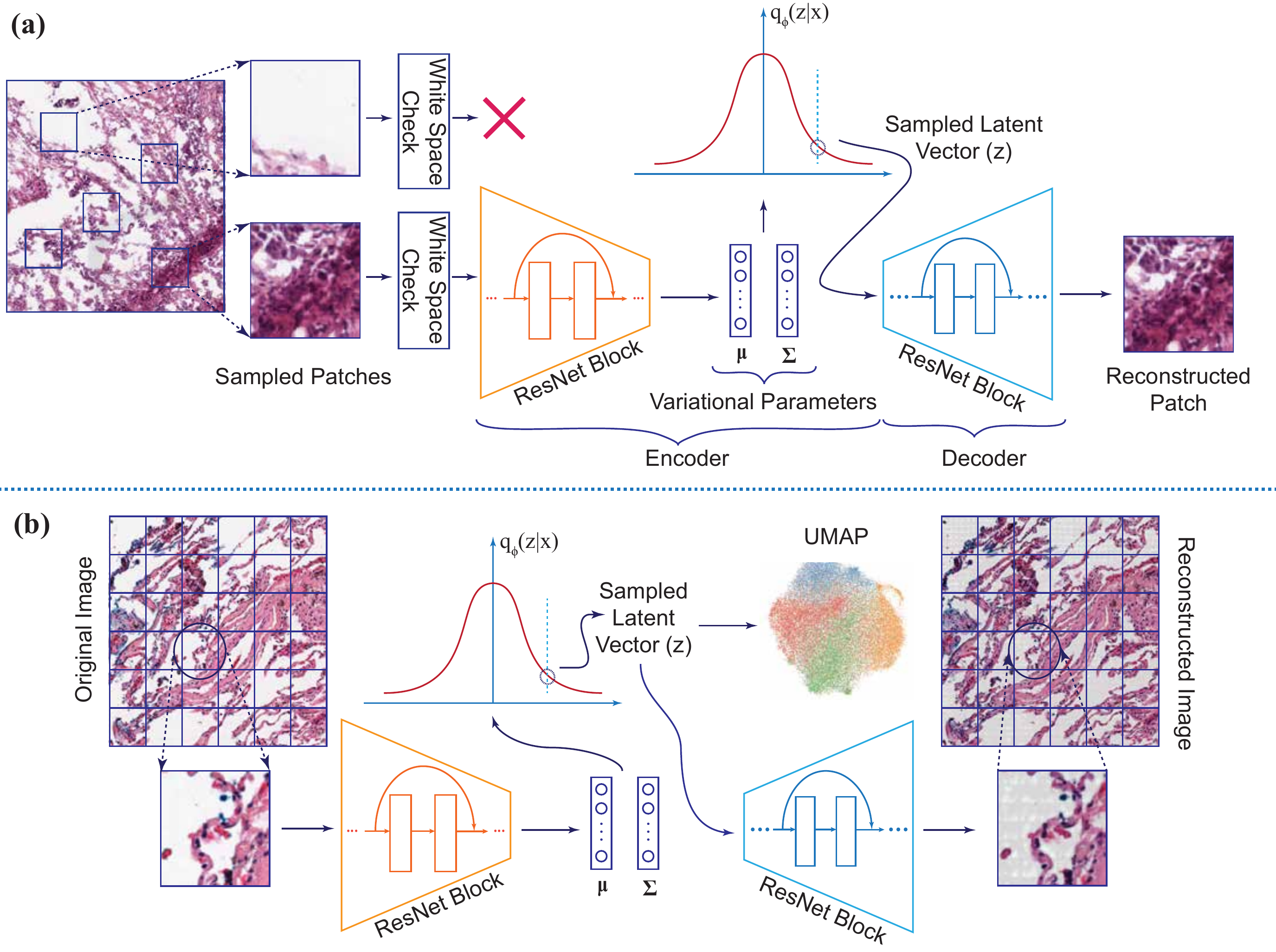}}
        \caption{\textbf{(a)} Overview of the VAE training pipeline. \textbf{(b)} Overview of the pipeline at inference. For generating UMAP plots, a similar patch sampling as training is used.}
        \label{fig:1}
    \end{figure}
    
    Histopathological images derived from cross sectional tissue microscopy are used in the clinical setting for diagnosis of various diseases and conditions \cite{gurcan2009histopathological}. Hemotoxylin and Eosin (H\&E) staining, which introduce a contrast dye for the discernment of nuclear and cytoplasmic structures, has long been used to determine carcinomal regions of excised tissue from cancer patients \cite{he2012histology}. For this reason, databases of tumor patient slides, such as the NIH Genomic Data Commons (GDC), have been compiled for researchers to access tens of thousands of cancer patients' histopathological data. The GDC itself contains more than 30,000 Whole Slide Images (WSIs) which, with each slide representing over a billion pixels each, is stored on over 20 TB of data. Most purposes, from retrieval to transmission, local storage, and data analysis would benefit from efficient, indexable storage structures of this WSI data \cite{niazi2019pathological}.  This is especially applicable to image search algorithms for large whole slide image databases \cite{kalra2020yottixel}.
    
    Several solutions have been proposed for the efficient storage and indexing of cancer tissue image data. Classic compression formulas such as JPEG2000 can successfully reduce image size at a compression ratio of 32:1 before becoming unusable for histopathological classification of malignancy \cite{krupinski2012compressing}. Compression and scaling has also been found to adversely effect tissue segmentation up to ratios of 50:1 \cite{konsti2012effect}. In contrast to discrete cosine transformation models, neural networks have been proven to retain high efficiency and fidelity in the lossy compression of image data \cite{soliman2006neural}. While neural networks seek to store image data in latent space representations, not every network does this at equivalent efficiency or accuracy \cite{jamil2022learning}. Several studies have demonstrated that Variational Auto Encoders (VAEs) retain higher image quality and lower noise ratios at extreme compression ratios \cite{hu2021learning,lombardo2019deep,yilmaz2021self}. Tellez et al., \cite{tellez2019neural}  showed in a benchmark study that VAE compression of medical tissue images to a latent space of 128 ($>$5000:1 compression ratio) retained the most details of the original whole slide image compared to 4 other encoders. In the current study, we develop a VAE to compress and index images in latent space for fast complex search of whole slide H\&E cancer images.
    
    \begin{figure}[t]
        \centering
        \centerline{\includegraphics[width=\linewidth]{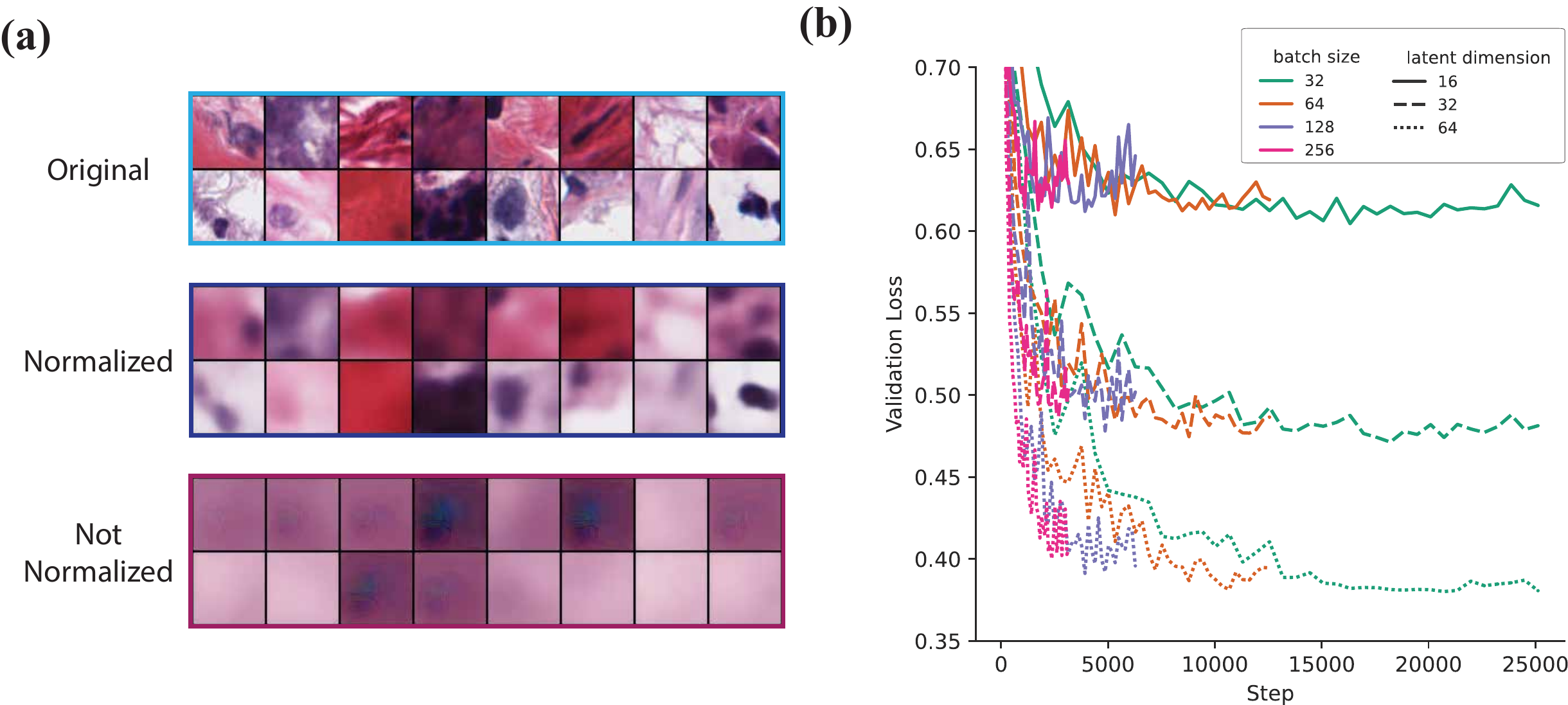}}
        \caption{\textbf{(a)} Example of how normalization affects the performance of our pipeline. Both models are trained using the exact same hyper-parameters (\texttt{latent\_dim} = 64). \textbf{(b)} The effect of batch size and latent dimension of validation loss. For better visualization, early stopping is not used for these experiments.}
        \label{fig:2}
    \end{figure}

    \section{Methods}
\label{sec:methods}

    \subsection{Dataset}
    \label{subsec:dataset}
    
    The dataset we used for this study is publicly available at the NCI GDC data portal (Sec. \ref{sec:availability}). These are real samples from cancer patients in the US, and all samples contain cancerous cells. For this study, We downloaded 20\% of the available \texttt{.svs} samples for primary sites: "Brain", "Breast", "Bronchus and Lung", and "Colon" (647, 551, 580, and 267 images, respectively).

    \subsection{Latent Variables and VAE}
    \label{subsec:vae}
    
        For an observation $\boldsymbol{x^{(i)}}$, its latent vector of variables is assumed to be an unobserved random variable $\boldsymbol{z^{(i)}}$ sampled from a lower dimension space (latent space) that is involved in producing $\boldsymbol{x^{i}}$ in a random process \cite{kingma_auto-encoding_2014}. For a particular task, it is assumed that using latent variables removes non-informative dimensionality and is suitable for downstream machine learning tasks. Since the latent space is unobserved, latent variables should be somehow inferred. Autoencoders and Variational Auto Encoders are two very effective methods for inferring these latent variables and encoding very high-dimensional data into highly compressed latent space with minimal loss of information. VAEs, as opposed to regular Auto Encoders, try to find a distribution for each latent variable, rather than single point estimate, resulting in a regularized latent space with generative capability.
        
        VAEs are comprised of two parts: an encoder and a decoder (Fig.\ref{fig:1}-a). If the latent variable $\boldsymbol{z^{(i)}}$ and data point $\boldsymbol{x^{(i)}}$ are sampled from parametric probability distributions $p_{\boldsymbol{\theta}}(\boldsymbol{z})$ and $p_{\boldsymbol{\theta}}(\boldsymbol{x} | \boldsymbol{z})$ for some parameter $\boldsymbol{\theta}$, then the encoder will try to estimate the approximate posterior $q_{\boldsymbol{\phi}}(\boldsymbol{z} | \boldsymbol{x})$ with variational parameter $\boldsymbol{\phi}$. The decoder tries to find the likelihood $p_{\boldsymbol{\theta}}(\boldsymbol{x} | \boldsymbol{z})$. The model can be trained by minimizing the loss introduced in Eq.\ref{eq:1} over all observations (\cite{kingma_auto-encoding_2014}). The first term of the loss is called the Kullback–Leibler (KL) divergence term, which is introduced to ensure that the variational approximation is as informative as the generative true posterior. The second term is reconstruction loss, which makes sure the generated output from the learned latent distribution is close to the original input. In  our experiments, we used a weighted loss with a KL term coefficient of $0.1$.
        
        \begin{equation}
        \begin{aligned}
            \mathcal{L} \left(\boldsymbol\theta, \boldsymbol\phi; \boldsymbol{x^{(i)}} \right) =& -D_{KL} \left( q_{\boldsymbol\phi} \left( \boldsymbol{z}|\boldsymbol{x^{(i)}} || p_{\boldsymbol\theta} \left( \boldsymbol{z} \right)\right) \right)\\
            & + \mathbb{E}_{q_{\boldsymbol\phi} \left( \boldsymbol{z}|\boldsymbol{x^{(i)}} \right)} \left[ \log{p_\theta \left( \boldsymbol{x^{(i)}} | \boldsymbol{z} \right)} \right]
        \end{aligned}
        \label{eq:1}
        \end{equation}

    \subsection{Training and Inference Pipelines}
    \label{subsec:pipelines}
    
        As illustrated in Fig.\ref{fig:1}, we use two pipelines for training and inference. For the training phase (Fig.\ref{fig:1}-a), a selected number of patches from whole slide images (WSIs) in the training and validation set are randomly sampled. A white space filter is utilized to ensure that these patches are not blank, and that they do not overlap. The mean and standard deviation of all patches sampled from the training set is calculated and all patches are normalized using the standard score method with these values (not shown in Fig.\ref{fig:1}). The inverse transformations are also stored to be applied to the outputs of the model.
        
       Our model assumes a Gaussian prior and a Gaussian approximate posterior. The encoder learns the parameters of the Gaussian prior and the decoder uses a re-parameterized sample from this prior and tries to reconstruct the input. Both encoder and decoder use ResNet18 (\cite{he2016deep}) architectures. A ResNet50 archiecture (not-shown) provides similar performance; ResNet18 was selected to keep the number of model parameters as small as possible for future downstream deployment in the clinic. 
        
        During inference (Fig.\ref{fig:1}-b), to perform a  a compression/decompression task, the test image is fully tiled. Each patch is then fed into the trained networks and stitched together once all patches are reconstructed. However, for the UMAP experiment, the same patch sampling algorithm used for training is used to generate random patches to be fed to the model. For the reconstruction task, we want a whole image, but for the UMAP plot, sample latent variables are enough.
        
    \subsection{Dimension Reduction and UMAP}
    \label{subsec:umap}
    
        Uniform Manifold Approximation and Projection (\cite{mcinnes2018umap}) is a manifold based dimensionality reduction algorithm used for visualizing and clustering high dimensional datasets. This algorithm tries to reduce the points in a manner that the distance between resulting points would be still meaningful. UMAP is utilized to visualize and demonstrate that not only do the latent vectors learned by our pipeline provide visually accurate decompressed images, but also they contain relevant clinical information from different cancer types (Fig.\ref{fig:4}). UMAP can use many metrics for distance calculation; "cosine similarity" was selected for its ability to capture correlation features.
        
        \begin{figure}[t]
            \centering
            \centerline{\includegraphics[width=\linewidth]{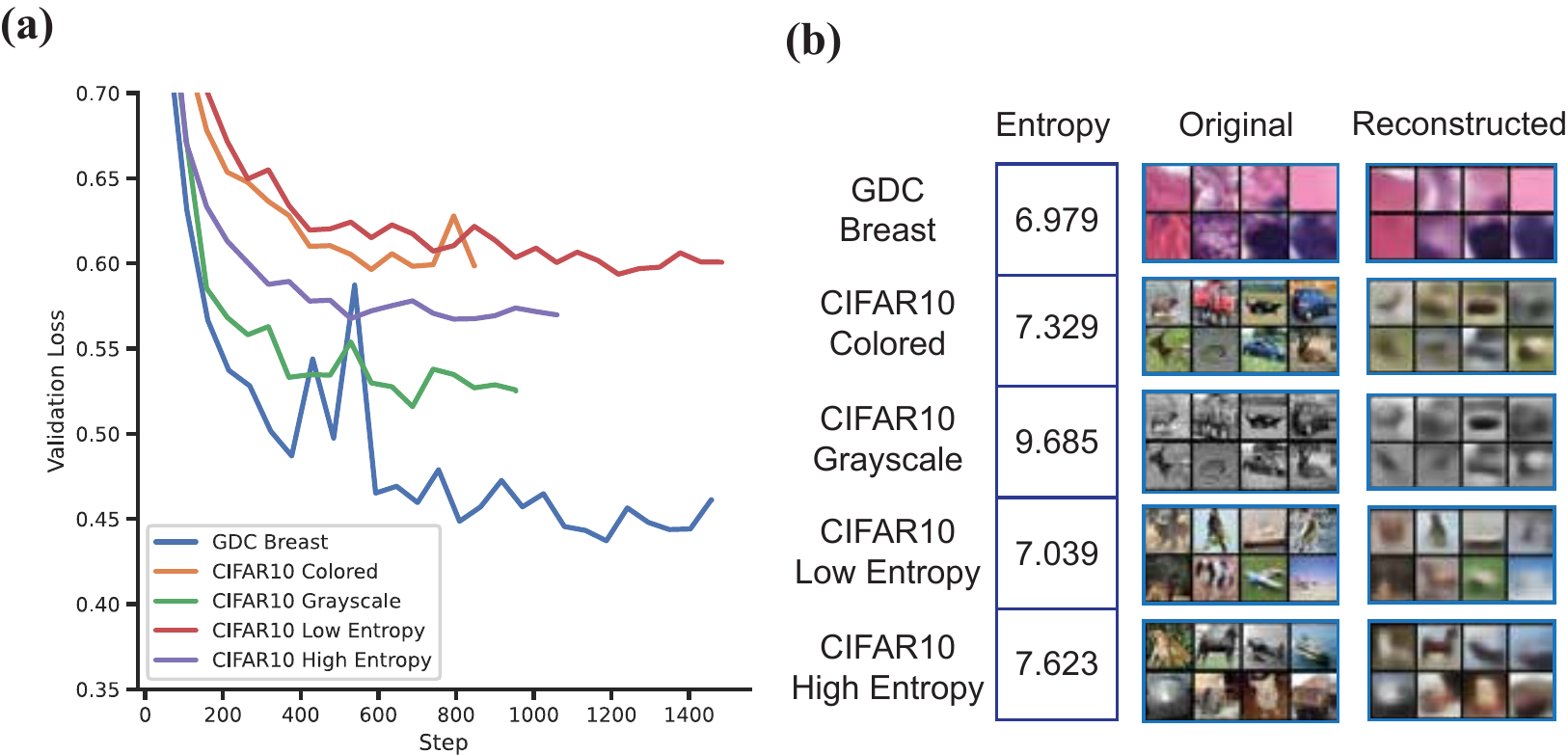}}
            \caption{Effect of dataset entropy and color content on peroformance. All hyper-parameters are the same for all 5 models.}
            \label{fig:3}
        \end{figure}
    \section{Settings and Experiments}
\label{sec:experiments}
    
    In this section, we summarize different scenarios and their experimental settings used for training and validating the compression and latent space approximation of histopathology images, and establish that the compression ratio our pipeline achieves is state of the art.
    
    \subsection{Training Settings}
    \label{subsec:training}
    
        For hyperparamter tuning, the effect of normalization of data on the quality of outcome was tested (based on visual inspection), and it was concluded that normalization is necessary for acceptable results (Fig. \ref{fig:2}-a). Since all datasets are normalized using the same procedure, the validation can be perceived as a metric to compare the performance of different models on different datasets. All experiments are conducted using and early stopping on validation loss with \texttt{patience} = 5 unless mentioned otherwise.
        
        As illustrated in Fig. \ref{fig:2}-b, higher batch sizes result into faster objective minimization, but lower batch sizes eventually results in better validation loss due to a higher regularization effect (\cite{Goodfellow-et-al-2016}). To take the middle ground, all experiments were conducted using a batch size of 128 unless mentioned otherwise. Also, as expected, higher latent dimensions resulted into a better performance.
        
        The model is developed with PyTorch Lightning API. All experiments were conducted using the DDP parallelization strategy on an NVIDIA DGX A100 with 8, 80 GB A100 GPUs, and a learning rate of $10^{-4}$. 
    
    \subsection{Compression Experiments}
    \label{subsec:compression}
    
        Experimental results demonstrate a better performance of our compression model on histopathology slides than is achieved on images of every day objects datasets such as in CIFAR10 (\cite{krizhevsky2009learning}). We first hypothesised that this diffeence is rooted in the difference of entropy between the average image in these two datasets. Entropy is a way of calculating the context information of a datapoint. We reasoned that low entropy images are more compressible  han high entropy ones. Therefore, we divided the CIFAR10 dataset by entropy with a high entropy fold (average entropy = 7.623) and a low entropy fold (average entropy = 7.039), each containing 30,000 images, and ran two experiments to see which one is more compressible when fed through our model. For both experiments, batch size was set to 256, latent dimension was set to 16, and the input images were of dimension $ 32 \times 32 \times 3 $. The results are shown in Fig. \ref{fig:3}. The final validation loss for low entropy and high entropy datasets are 0.601 and 0.570, respectively contradicted our original hypothesis. We ran the same experiment on the same number of patches sampled for the breast cancer slides, and although having lower entropy, it showed a better performance (numbers are reported in Fig\ref{fig:3}). Hence, we concluded that entropy is not a reliable factor to explain the SOTA performance of our VAE compression pipeline on cancer imaging data.
    
        We then hypothesized that color distribution may be a contributing factor. H\&E slides are limited to the colors present in tissue, while CIFAR10 images have a more diverse color distribution. For this hypothesis, we randomly chose 30,000 images from CIFAR10 dataset. Using the same settings, we ran one experiment on the sampled images and another on the same images but with grayscale transformation to eliminate olor diversity. The final validation loss for colored dataset is 0.599 and for the grayscale dataset is 0.525 (Fig. \ref{fig:2}-a). The lower validation loss indicates that less color content can be attributed to a better comprehensibility.

    \subsection{Validation Experiments}
    \label{subsec:val_exp}
    
        In order to examine whether the latent space preserves necessary information for downstream clinical tasks, we tested the accuracy of original slide images against regenerated slide images on CLAM (\cite{lu2021data}), the state-of-the-art model in lung cancer classification from H\&E slides. We first used CLAM on the original test set for the two classification tasks, i.e. "tumor vs. normal" and "sub-typing" between Lung Adenocarcinomas (LUAD) and Squamous Cell Carcinomas (LUSC). Then, we created a reconstructed (post compression) version of the test set using our inference pipeline (Fig. 1-b). This reconstructed test set was then run through the same classification problems as the original images. We then calculated the percentage of the images that had the same label for both original and reconstructed images over all test images as a measure of performance and observed that our compression did not decrease performance on clinical application tasks.
        
         To test the clinical information preservation of the latent space, we chose a model trained on lung tissues with the highest compression ratio (1:512) in our pipeline (Sec. \ref{subsec:val_exp}). This compression ratio is twice as high as the best models introduced in the literature (\cite{tellez2019neural, chen_fast_2022}). For the "tumor vs. normal" task, the reconstructed images did not show loss of performance, however, this level of compression made it difficult for lung cancer sub-typing model to perform as before.
        
        We used 900 images from the GDC TCGA (Sec. \ref{sec:availability}) including 450 samples for each LUAD and LUSC sub-types for training and 100 images (50 LUAD, 50 LUSC) for testing. The CLAM model has 10-fold validated pre-trained weights; thus, we calculated the performance in a 10-fold setting, too.
        
        \begin{figure}[t]
            \centering
            \begin{minipage}[b]{0.48\linewidth}
                \centering
                \centerline{\includegraphics[width=1\linewidth]{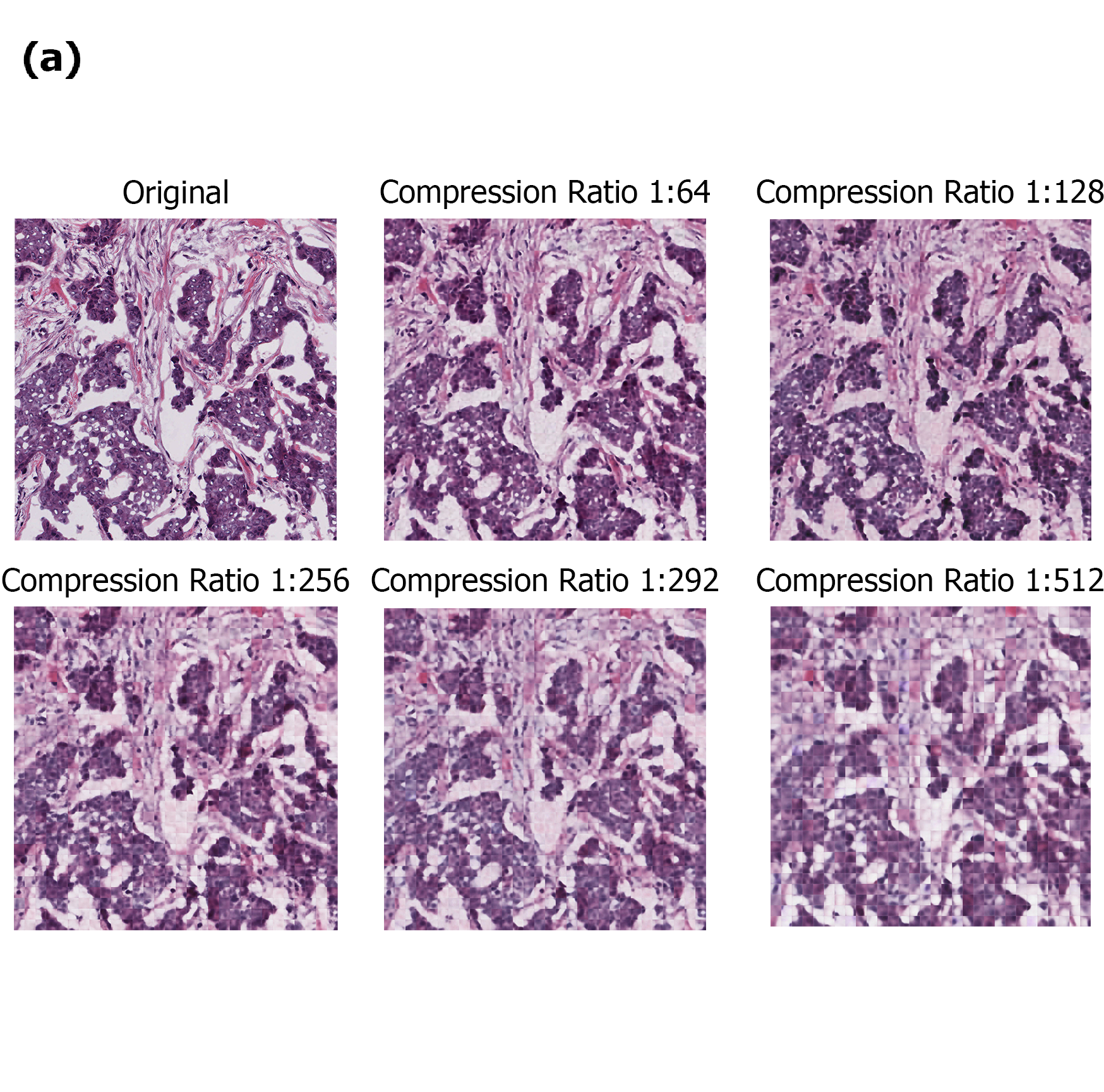}}
            \end{minipage}
            \begin{minipage}[b]{0.5\linewidth}
                \centering
                \centerline{\includegraphics[width=\linewidth]{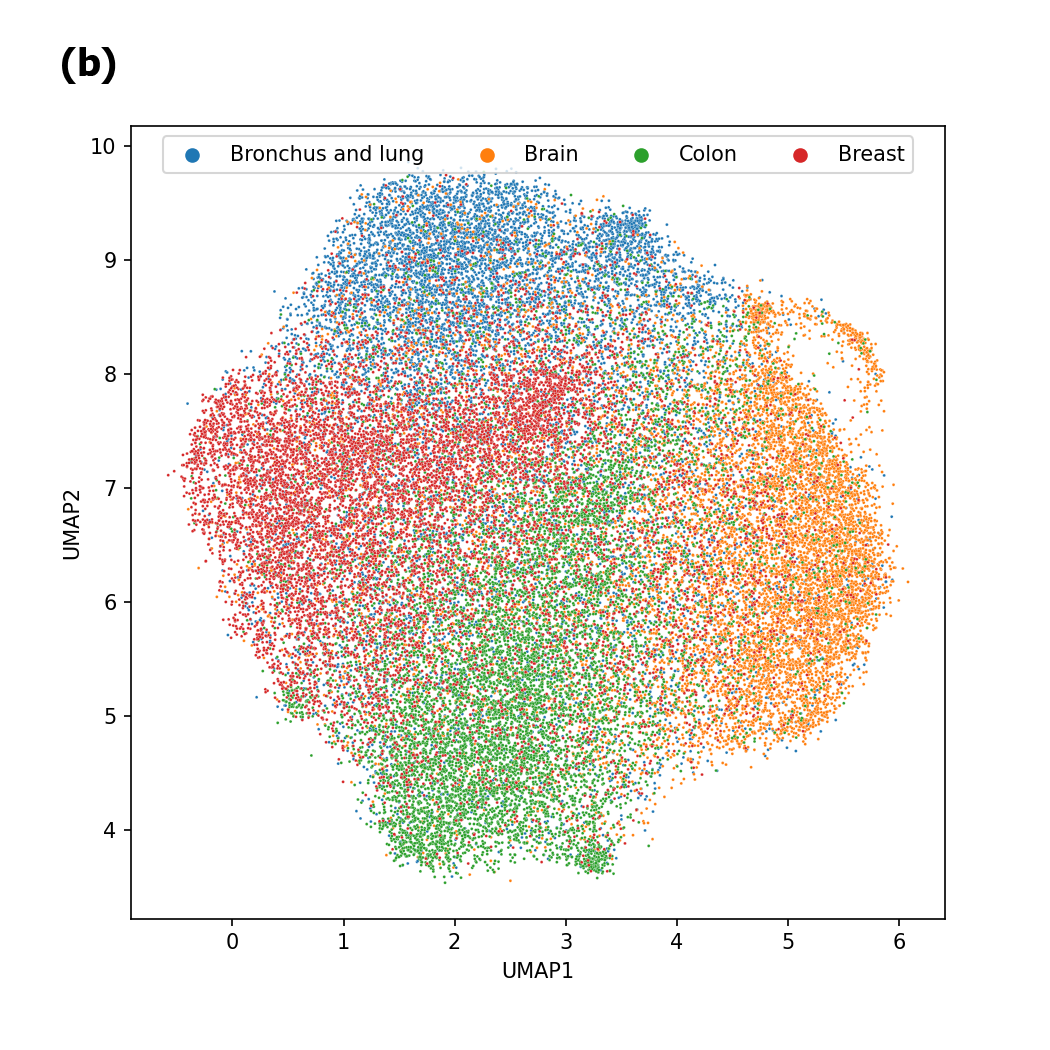}}
            \end{minipage}
            \caption{\textbf{(a)} The reconstruction results for breast cancer tissues at 5 different compression ratios. \textbf{(b)} UMAP plot generated on 4 different tissue types with a compression ratio of 1:64.}
            \label{fig:4}
        \end{figure}

    \subsection{UMAP Experiments}
    \label{subsec:umap_exp}
    
        To show that the latent space preserves important and clinically relevant information, 4 models were trained with a latent space of size 64 on different tissue types (brain, breast, bronchus and lung, and colon) on 20,000 patches of size $64 \times 64$ pixels, and tested them on 10,000 patches from their respective tissue type. We then ran the latent vectors of the test patches through the UMAP algorithm using the "cosine" distance as the similarity metric. The results are shown in Fig.\ref{fig:4}. 

    \section{Results and Conclusion}
\label{sec:conclusion}

    Fig. \ref{fig:4}-a shows the impact of various compression ratios on VAE output images. At lower compression ratios, reconstructed images more closely resemble original input images. Importantly, we see a marked improvement in histologic features that are critical for interpretability such as refined cell-to-cell borders and sharper demarcation of cytoplasmic vs. nuclei compartments. Moreover, in Fig. \ref{fig:4}-b, we use UMAP to visualize the latent space vectors learned by our pipeline. The UMAP captures intra-tumor and across-tumor relationships, separating all four tissue types into distinct clusters. Interestingly, clusters of brain and colon cancers share overlapping boundaries whereas the breast cancer cluster is uniquely separated from the brain cancer cluster. Also, the UMAP identifies a distinct sub-cluster of brain tumor samples that does not overlap with any other cancer types. 

     We envision our pipeline being useful to clinicians and researchers across multiple domains. One potential application is more accurate sub-typing and diagnoses of poorly understood cancers. A notable example of this is brain cancer, which contains over 150 different histologic subtypes, many of which are so rare that a pathologist may only encounter a handful of cases in his or her career (\cite{roetzer-pejrimovsky_digital_2022}). In our UMAP visualization of the latent space, there is an unexpected but distinct sub-cluster of brain tumor samples that does not overlap with other cancer types (Fig. \ref{fig:4}-b). Further characterization of this sub-cluster and its unique attributes could provide novel insights into intra-tumor relationships in brain cancer. 

    Our pipeline also facilitates experiments across different tumor types. The latent space separates breast, colon, lung/bronchus, and brain tissue into unique clusters, demonstrating the preservation of important histological features. Interestingly, we see a closer clustering between brain and colon cancer versus brain and breast or lung (Fig. \ref{fig:4}-b). More investigation into these relationships is warranted – one possible explanation of this phenomenon could be due to both brain and colon tissue containing ganglion nerve cells whereas breast and lung tissue do not. In the future, our embedding approach could be deployed to a hospital system and linked to the electronic health record (EHR) to help clinicians diagnose patients with rare disorders: the images closest to that of the input patient in UMAP embedding have the most similarities, and their records could be retrieved to better contextualize a differential diagnosis for the query patient. 
    
    However, our pipeline carries several limitations. To start, we will need to further explore acceptable thresholds of reconstruction loss introduced via our VAE-based architecture. Additionally, our model architecture lacks human interpretable features, which may lead to higher levels of end-user distrust as “peeking under the hood” to audit our model for biases or errors may be more limited. Along these lines, any insights or novel conclusions will still require manual review and interpretation by human pathologists. In future iterations of this work, we intend to improve upon these areas.

    \section{Data and Code Availability}
\label{sec:availability}

    All dataset used in this study are accessible from NCI GDC portal at \href{https://portal.gdc.cancer.gov/repository}{portal.gdc.cancer.gov/repository}. The code is also accessible at \href{https://github.com/jacobluber/uta_cancer_search}{github.com/jacobluber/uta\_cancer\_search}.
    \clearpage
    \section{Compliance with ethical standards}
This research study was conducted retrospectively using human subject data made available in open access by the Genomic Data Commons (GDC) provided by the National Cancer Institute of the National Institues of Health (NIH/NCI). Ethical approval was not required as confirmed by the license attached with the open access data.
    \section{Acknowledgments}
This work was supported by the Cancer Prevention and Research Institute of Texas (CPRIT) Recruitment of First-Time, Tenure-Track Faculty Members Grant (RR220015) (JML) and University of Texas System Rising STARs award (JML). The authors wish to thank Abhishek Dubey, Peng Jiang, , and Eytan Ruppin for fruitful discussions. The authors want to thank Thomas Shipman and Edward Gonzalez for their support of computing infrastructure. 
    
    
    \bibliographystyle{IEEEbib}
    \bibliography{refs}

\begin{thebibliography}{10}

\bibitem{gurcan2009histopathological}
Metin~N Gurcan, Laura~E Boucheron, et~al.,
\newblock ``Histopathological image analysis: A review,''
\newblock {\em IEEE reviews in biomedical engineering}, vol. 2, pp. 147--171,
  2009.

\bibitem{he2012histology}
Lei He, L~Rodney Long, et~al.,
\newblock ``Histology image analysis for carcinoma detection and grading,''
\newblock {\em Computer methods and programs in biomedicine}, vol. 107, no. 3,
  pp. 538--556, 2012.

\bibitem{niazi2019pathological}
M~Khalid~Khan Niazi, Yuzhang Lin, et~al.,
\newblock ``Pathological image compression for big data image analysis:
  Application to hotspot detection in breast cancer,''
\newblock {\em Artificial intelligence in medicine}, vol. 95, pp. 82--87, 2019.

\bibitem{kalra2020yottixel}
Shivam Kalra, Hamid~R Tizhoosh, et~al.,
\newblock ``Yottixel--an image search engine for large archives of
  histopathology whole slide images,''
\newblock {\em Medical Image Analysis}, vol. 65, pp. 101757, 2020.

\bibitem{krupinski2012compressing}
Elizabeth~A Krupinski, Jeffrey~P Johnson, et~al.,
\newblock ``Compressing pathology whole-slide images using a human and model
  observer evaluation,''
\newblock {\em Journal of pathology informatics}, vol. 3, no. 1, pp. 17, 2012.

\bibitem{konsti2012effect}
Juho Konsti, Mikael Lundin, et~al.,
\newblock ``Effect of image compression and scaling on automated scoring of
  immunohistochemical stainings and segmentation of tumor epithelium,''
\newblock {\em Diagnostic Pathology}, vol. 7, no. 1, pp. 1--9, 2012.

\bibitem{soliman2006neural}
Hamdy~S Soliman and Mohammed Omari,
\newblock ``A neural networks approach to image data compression,''
\newblock {\em Applied Soft Computing}, vol. 6, no. 3, pp. 258--271, 2006.

\bibitem{jamil2022learning}
Sonain Jamil, Md~Piran, et~al.,
\newblock ``Learning-driven lossy image compression; a comprehensive survey,''
\newblock {\em arXiv preprint arXiv:2201.09240}, 2022.

\bibitem{hu2021learning}
Yueyu Hu, Wenhan Yang, et~al.,
\newblock ``Learning end-to-end lossy image compression: A benchmark,''
\newblock {\em IEEE Transactions on Pattern Analysis and Machine Intelligence},
  2021.

\bibitem{lombardo2019deep}
Salvator Lombardo, Jun Han, et~al.,
\newblock ``Deep generative video compression,''
\newblock {\em Advances in Neural Information Processing Systems}, vol. 32,
  2019.

\bibitem{yilmaz2021self}
M~Ak{\'\i}n Y{\'\i}lmaz, Onur Keles{\c{s}}, et~al.,
\newblock ``Self-organized variational autoencoders (self-vae) for learned
  image compression,''
\newblock in {\em 2021 IEEE International Conference on Image Processing
  (ICIP)}. IEEE, 2021, pp. 3732--3736.

\bibitem{tellez2019neural}
David Tellez, Geert Litjens, et~al.,
\newblock ``Neural image compression for gigapixel histopathology image
  analysis,''
\newblock {\em IEEE transactions on pattern analysis and machine intelligence},
  vol. 43, no. 2, pp. 567--578, 2019.

\bibitem{kingma_auto-encoding_2014}
Diederik~P. Kingma and Max Welling,
\newblock ``Auto-{Encoding} {Variational} {Bayes},'' May 2014,
\newblock arXiv:1312.6114 [cs, stat].

\bibitem{he2016deep}
Kaiming He, Xiangyu Zhang, et~al.,
\newblock ``Deep residual learning for image recognition,''
\newblock in {\em Proceedings of the IEEE conference on computer vision and
  pattern recognition}, 2016, pp. 770--778.

\bibitem{mcinnes2018umap}
Leland McInnes, John Healy, and James Melville,
\newblock ``Umap: Uniform manifold approximation and projection for dimension
  reduction,''
\newblock {\em arXiv preprint arXiv:1802.03426}, 2018.

\bibitem{Goodfellow-et-al-2016}
Ian Goodfellow, Yoshua Bengio, and Aaron Courville,
\newblock {\em Deep Learning},
\newblock MIT Press, 2016,
\newblock \url{http://www.deeplearningbook.org}.

\bibitem{krizhevsky2009learning}
Alex Krizhevsky, Geoffrey Hinton, et~al.,
\newblock ``Learning multiple layers of features from tiny images,''
\newblock 2009.

\bibitem{lu2021data}
Ming~Y Lu, Drew~FK Williamson, et~al.,
\newblock ``Data-efficient and weakly supervised computational pathology on
  whole-slide images,''
\newblock {\em Nature Biomedical Engineering}, vol. 5, no. 6, pp. 555--570,
  2021.

\bibitem{chen_fast_2022}
Chengkuan Chen, Ming~Y. Lu, et~al.,
\newblock ``Fast and scalable search of whole-slide images via self-supervised
  deep learning,''
\newblock {\em Nature Biomedical Engineering}, pp. 1--15, Oct. 2022,
\newblock Publisher: Nature Publishing Group.

\bibitem{roetzer-pejrimovsky_digital_2022}
Thomas Roetzer-Pejrimovsky, Anna-Christina Moser, et~al.,
\newblock ``The {Digital} {Brain} {Tumour} {Atlas}, an open histopathology
  resource,''
\newblock {\em Scientific Data}, vol. 9, no. 1, pp. 55, Feb. 2022.

\end{thebibliography}

\end{document}